\documentclass[prl,twocolumn,showpacs,preprintnumbers,amsmath,amssymb,superscriptaddress]{revtex4}

\usepackage{graphicx}
\usepackage{dcolumn}
\usepackage{color}

\allowdisplaybreaks
\newcommand{\beq}{\begin{equation}}
\newcommand{\eeq}{\end{equation}}
\newcommand{\beqa}{\begin{eqnarray}}
\newcommand{\eeqa}{\end{eqnarray}}

\begin{document}
\title{Isotope effect in the pseudogap of high-temperature superconducting copper oxides}

\author{G.~Sangiovanni}
\affiliation{Institut f\"ur Festk\"orperphysik, TU Wien, Vienna, Austria}

\author{O.~Gunnarsson}
\affiliation{Max-Planck Institut f\"ur Festk\"orperforschung, Heisenbergstr. 1, D-70569 Stuttgart, Germany}


\begin{abstract}
We study cuprates within Dynamical Cluster Approximation and find the pseudogap displays an isotope effect of the same sign as observed experimentally. Notwithstanding the non-phononic origin of the pseudogap the interplay between electronic repulsion and retarded phonon-mediated attraction gives rise to an isotope dependence of the antinodal spectra. Due to the strong momentum differentiation, such interplay is highly non-trivial and leads to the simultaneous presence of heavier quasiparticles along the nodal direction. We predict an isotope effect in electron-doped materials.
\end{abstract}
\date{\today}
\maketitle

Copper oxide superconductors are characterized by the highest superconducting transition temperatures ever measured and, twenty-five years after their discovery the pairing mechanism has not yet been clarified. 
However, this is probably not the main reason why these materials are so intriguing. 
Indeed, due to the complex interplay between their electronic, magnetic and lattice degrees of freedom we still have a very incomplete understanding even of their normal (non-superconducting) phase. 
The most well-known anomaly characterizing the cuprates at small values of doping is the presence of a pseudogap. 
Using the language of angular resolved photoemission spectroscopy, the pseudogap is a strong suppression of the photoemission signal in selected parts of the Brillouin zone (BZ), occurring below a characteristic temperature called $T^*$.
This phenomenon is believed to be so relevant that any complete theory of cuprates should explain it.

Differently from the superconducting transition temperature $T_c$, $T^*$ displays a quite marked isotope effect characterized by a negative coefficient, i.e. it increases upon substituting Oxygen or Copper atoms with correspondingly heavier isotopes \cite{temprano,tempranoPRB66,tempranoEPJB,isotope8}.
This represents an unusual dependence, having conventional BCS theory in mind.
Some possible scenarios for its explanation have been proposed:
Assuming that the pairing in cuprates is directly mediated by phonons \cite{ranninger} so that the pseudogap is associated to the binding of two polarons with no long-range phase coherence, Ranninger calculated the isotope shift of $T^*$ and found good qualitative agreement with experiments. 
The main drawback of this approach is that a small electronic repulsion is enough to make bipolaron formation energetically highly unfavorable, and in cuprates local Coulomb repulsion can by no means be neglected. 
Another theoretical prediction available is based on the idea of a quantum critical point at optimal doping: By analyzing the effects of critical fluctuations Andergassen {\it et al.} proposed a consistent picture of the experimental observation \cite{sabine1}.
Yet, no calculation has been done so far to see whether or not an isotope effect on $T^*$ is to be expected within probably the simplest model for cuprates, namely a one-band tight-binding model for electrons or holes on the Copper sites with nearest-neighbor and diagonal hopping $t$ and $t'$ respectively, experiencing a local Hubbard repulsion $U$ when occupying the same lattice site. 
Such model yields indeed one of the most successful descriptions of cuprates and many of the experimental observations such as $d$-wave superconductivity and the momentum differentiation characterizing the pseudogap have been qualitatively fairly well reproduced when solving it within Dynamical Cluster Approximation (DCA), Cluster Dynamical Mean Field Theory (CDMFT) and similar approaches \cite{DCA1,cellular1,review05,gull_mom-diff,aichhornPRB74}. 

\begin{figure}[ht]
\begin{center}
\includegraphics[width=8.5cm]{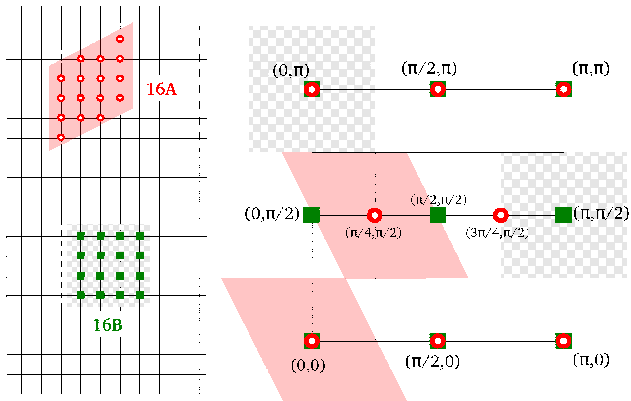}
\end{center}
\caption{(Color online) (Left) 16-site clusters considered (the notation is the same as in Ref. \cite{maierPRL95}). (Right) Corresponding cluster momenta ${\bf K_c}$ in the BZ. The ${\bf K_c}$ marked by both the green full square and the red open circle are present in both geometries. The reddish and checkerboard gray areas denote the coarse graining regions for 16A and B respectively.}
\label{fig1}
\end{figure}

Yet, recent theoretical studies showed that in order to get a satisfactory quantitative agreement with photoemission and optics experiments the Hubbard (or the $t$-$J$) model is in fact not enough. 
Several experiments \cite{kink2,kmshen,pint3,pint7,pint6} suggest a strong coupling to selected phonon modes and indeed things improve substantially if a coupling to phonons is included \cite{ciuchi-tJ,mishchenko,olle-polaron,StJ3,cappellutiPRB79,mishchenkoPRL100,mishchenko_rev2}.
The size and the symmetry of such coupling have been previously calculated by combining \emph{ab-initio} calculations and many-body effects \cite{pint4,olle1,threeband,zhangberkley,communication}.
Therefore we solve here the Hubbard model in the presence of electron-phonon ($e$-ph) coupling within DCA on 16-site clusters, as shown in Fig. \ref{fig1}.
We focus on its effects on the spectral function and look for the presence of an isotope effect in the pseudogap.

Previous DCA studies of small clusters indicated that the pseudogap is very little affected and superconductivity gets rapidly suppressed upon increasing the $e$-ph coupling \cite{macridin,macridin_isot}.
One may therefore expect our study to reveal hardly any isotope effect on the pseudogap 
This property in DCA is indeed a genuine electronic correlation effect and it is present already in the absence of phonons. 
On the other hand the different degrees of freedom are mutually entangled in cuprates so that even quantities which have no direct phononic origin may display an isotope dependence.
Our study gives the novel evidence that this is indeed the case for the hole- and electron-doped ($e$-doped) cuprates: The Hubbard model has a pseudogap already in the absence of phonons and an $e$-ph coupling gives only a small change in its size and its momentum dependence.
Nevertheless the pseudogap becomes isotope dependent consistently, as our results show, with experiments.  
The consequence of this is important as the common way of looking at the isotope effect in cuprates is still based on the idea that either this is an extrinsic effect with little connection with the actual physical mechanisms or that it is a consequence of theories in which pairing is phonon-mediated, thus hardly realizable in cuprates.
Our study shows that both of the above arguments are false: an isotope effect on $T^*$ is an intrinsic effect arising when a strong local Coulomb repulsion has to live together with a retarded attraction mediated by phonons and it is possible to describe the experimental data without necessarily invoking doubtful phononic pairing mechanisms \cite{lee_isot}. 

\begin{figure}[ht]
\begin{center}
\includegraphics[width=8.5cm]{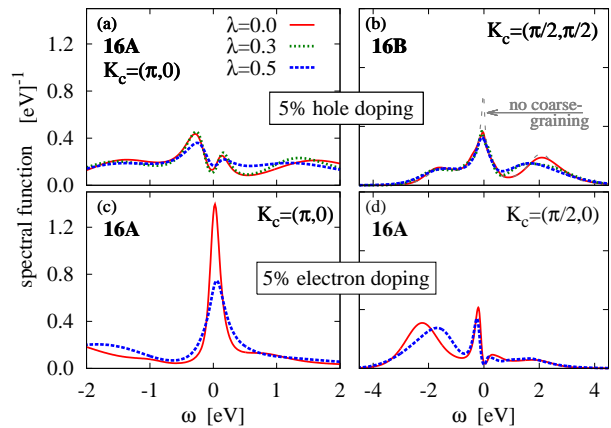}
\end{center}
\caption{(Color online) 16-site cluster spectra at $T\!\!=\!\!580$K for different cluster momenta ${\bf K_c}$ for hole- (a-b) and $e$-doping (c-d). $U\!\!=\!\!8|t|$, $|t|\!\!=\!\!400$m$eV$, $|t'|\!\!=\!\!120$m$eV$. $\omega_{ph}\!\!=\!\!200m$eV and $100m$eV for hole- and $e$-doped respectively. ${\bf K_c}$-resolved DCA spectra do not depend on any interpolation scheme. For the analytic calculation we have used Maximum Entropy \cite{jarrgub,maxent_split}.} 
\label{fig2}
\end{figure}
The first question is therefore ``What is the effect of phonons in the different sectors of the Brillouin zone?''.
The form of the $e$-ph coupling term we consider is derived from first-principle arguments following Ref. \onlinecite{olle1}. 
We show calculations for phonons of Holstein-type but we also considered breathing-type of modes in which the Oxygen atoms move in- and outwards the central Copper.
For the DCA solver we have used is a Hirsch-Fye Quantum Monte-Carlo \cite{HF}. 

In Fig. \ref{fig2} we show the spectral function for different cluster momenta ${\bf K_c}$ for 5\% hole- and $e$-doping. 
For hole-doping the pseudogap is in the ${\bf K_c}\!=\!(\pi,0)$ sector (panel a) while for $e$-doping the pseudogap is much smaller, though still somewhat visible at ${\bf K_c}\!=\!(\pi/2,0)$ (panel d) as well as at ${\bf K_c}\!=\!(\pi/4,\pi/2)$ \cite{senechalPRL92,kyungPRL93,macridinPseud,sordiPRL104,note_el-dop}.
It is clear how in both cases the changes to the pseudogap features induced by the $e$-ph coupling are rather small.
We observed that quantitatively such changes are fairly sensitive to the value of $t'$ used, the cluster size as well as whether the $e$-ph coupling is increased beyond to the values shown here. 
Our definition of the $e$-ph coupling ${\lambda}$ is the ratio of the bipolaronic energy to the electronic bandwidth given by $W\!\!=\!\!8|t|\!\!=\!\!3.2$eV and we consider $U\!\!=\!\!W$.
As we will see the point is that even if the $\lambda$-dependence of the pseudogap is small, this is enough to give an appreciable isotope effect (i.e. a dependence of the pseudogap on the ion's mass, in other words on $\omega_{ph}$). 

It is interesting to analyze the effect of the $e$-ph coupling on quasiparticles.  
For hole-doping these are in the ${\bf K_c}\!\!=\!\!(\pi/2,\pi/2)$ sector (Fig. \ref{fig2}b) while for $e$-doping these are at ${\bf K_c}\!\!=\!\!(\pi,0)$ (Fig. \ref{fig2}c).
At the temperature considered here (580K) DCA gives in the hole-doped region much broader quasiparticles than for $e$-doping.
The natural consequence of this is that we can observe effects of the $e$-ph coupling in the $e$-doped side much better. 
Indeed in Fig. \ref{fig2}c we see a substantial suppression of spectral weight (about a factor 2) while quasiparticles of panel b (hole-doping) are quite insensitive to $\lambda$.
The insensitiveness of hole-doped quasiparticles to $\lambda$ is found also if we look at spectral functions without ``coarse-graining'', i.e. without the average over the BZ patches characterizing DCA \cite{review05} (sketched in Fig.\ref{fig1} as transparent regions).
This procedure makes quasiparicle sharper, as indicated (for $\lambda\!\!=\!\!0$ only) by the arrow in Fig. \ref{fig2}b.
Yet, this is not enough to make $e$-ph effects discernible on the hole-doped side. 
Lower temperature studies with continuous-time QMC solvers (see Ref. \onlinecite{gullRMP} for a review) are underway to understand whether this is more than just a temperature effect.

\begin{figure}[ht]
\begin{center}
\includegraphics[width=8.5cm]{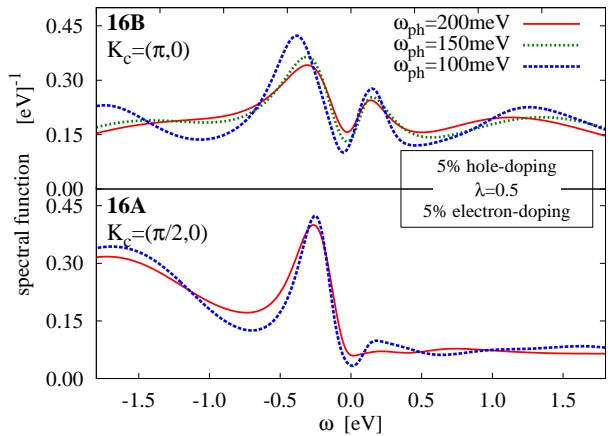}
\end{center}
\caption{(Color online) Pseudogap for different values of $\omega_{ph}$ for 5\% hole- and $e$-doping, upper and lower panel respectively.}
\label{fig3}
\end{figure}
\begin{figure}[ht]
\begin{center}
\includegraphics[width=8.5cm]{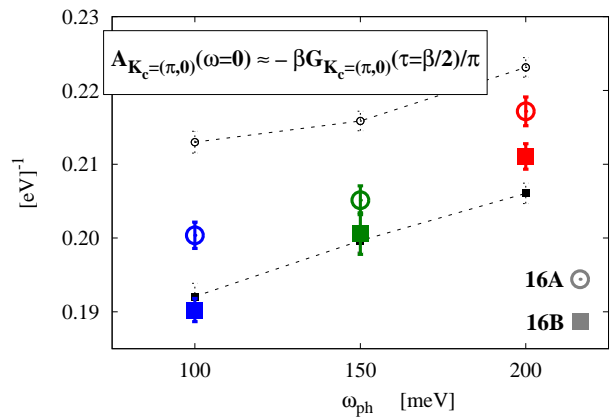}
\end{center}
\caption{(Color online) $A_{{\bf K_c}\!=\!(\pi,0)}(\omega\!\!=\!\!0)$ for 5\% hole doping for both clusters. The dashed black lines show the same quantity obtained for $\lambda\!\!=\!\!0$ and changing $U$ ($\omega_{ph}\!\!=$100, 150 and 200m$eV$ corresponds to $U\!=$3.00, 2.95 and 2.90$eV$ respectively).}
\label{fig4}
\end{figure}

The second question is then ``Does the $e$-ph interaction lead to an isotope effect in the pseudogap consistent with experiments?''. 
The answer can be directly read off from Fig. \ref{fig3} and \ref{fig4}.
In Fig. \ref{fig3} the spectral functions at 5\% hole- and $e$-doping is shown for the ${\bf K_c}$-sectors where the pseudogap is observed.
The outcome of DCA is clear: the pseudogap increases upon decreasing $\omega_{ph}$, i.e. upon increasing the ions' mass. 
This is the same trend found in experiments on hole-doped cuprates in which $T^*$ increases upon substituting $^{16}$O with $^{18}$O \cite{temprano,tempranoPRB66,tempranoEPJB,isotope8}. 

In order to extract more quantitative information from our data we look at a quantity which, unlike the spectral function, is independent on the analytical continuation method.
This is the value of the ${\bf K_c}$-resolved Green's function at imaginary time $\tau\!\!=\!\!\beta/2$. 
Indeed, this quantity does not depend on Maxent and it gives an estimate of the spectral weight at ${\bf K_c}$ around $\omega\!\!=\!\!0$, $A_{\bf K_c}(\omega)$.
The results for different values of $\omega_{ph}$ are shown by the bigger symbols in Fig. \ref{fig4}.
For both the 16A and B cluster geometry $A_{{\bf K_c}\!=\!(\pi,0)}(0)$ decreases as $\omega_{ph}$ decreases, as we expected from Fig. \ref{fig3}.
Even though we considered a pretty coarse grid in $\omega_{ph}$, from the slope of these curves we can extract an estimate of the isotope coefficient on $A_{{\bf K_c}\!=\!(\pi,0)}(0)$ for 5\% hole-doping. 
The value we get is about 0.1.
As it must be, the sign is opposite to that of the isotope coefficient on the pseudogap amplitude, because a larger $A_{{\bf K_c}\!=\!(\pi,0)}(0)$ indicates a less pronounced pseudogap.
Our study also allows us to make a prediction for what should happen on the $e$-doped side for which, to our knowledge, no isotope effect experiments have been performed hitherto. 
We expect them to reveal a smaller isotope effect in the pseudogap (if visible) compared to the hole-doped side, but a substantial $e$-ph effect on quasiparticles. 
This is different from what predicted by the resonant pairing scenario \cite{ranninger}, in which the isotope coefficient is essentially doping independent. 

More than one physical argument can be given to interpret our DCA results:
Firstly, the trend of an increase in the pseudogap upon decreasing $\omega_{ph}$ can be qualitatively seen in terms of a simple parametrization of the self-energy in terms of a pseudogap contribution plus a standard $e$-ph second-order diagram. 
An effect which may also play a role is a ph-induced broadening of the higher-energy features in the spectrum which gets smaller upon decreasing $\omega_{ph}$ thus increasing the pseudogap \cite{fratiniPRB72}.
Furthermore, our findings may be related to a recently proposed composite-fermion theory in which a reduction of the degree of coherence of quasiparticles due a smaller $\omega_{ph}$ would lead to an increase of the ``cofermion'' binding energy and consequently of the pseudogap \cite{cofermion}. 
Here we however want to discuss the following very simple interpretation: 
Phonons mediate an attraction between electrons with a characteristic frequency scale set by $\omega_{ph}$.
This introduces a reduction of the degree of correlation which is intrinsically $\omega_{ph}$-dependent, a fact that can naturally give an isotope effect on quantities which depend on correlation rather than on phononic mechanisms \cite{StJ2}.
Our DCA results suggest that this mechanism can describe the isotope effect in the pseudogap of cuprates. 
This can be seen in Fig. \ref{fig4} where we plot $A_{{\bf K_c}\!=\!(\pi,0)}(0)$ obtained by just slightly rescaling $U$ keeping $\lambda \!\! = \!\! 0$ (small symbols connected by dashed lines). 
The reduction obtained by rescaling $U$ is fairly similar to that obtained by really changing $\omega_{ph}$. 
Of course in our 16-site DCA cluster it is possible to describe a much richer behavior beyond this simplified picture: we indeed observe the ph-induced reduction of $U$ \emph{together with} a sizable mass enhancement of coherent quasiparticle. 
In Fig. \ref{fig2} we not only see the tendency of the Hubbard band features to get closer together upon increasing $\lambda$ but we also see the quasiparticle weight suppression, describing the typical loss of coherence induced by phonons in a metallic system.
In other words we have a slightly less correlated system in the pseudogapped sectors of the BZ due to the ph-induced reduction of $U$ but heavier quasiparticles in the coherent sectors due to more standard $e$-ph effects. 
For a complete explanation of the isotope effect on the pseudogap it will be necessary to understand which correlations are directly affected by the $e$-ph coupling and, among these, which are the most important ones for the pseudogap formation. 
This remains an open question. 
We have shown that in the presence of a strong differentiation in momentum space, like in cuprates \cite{gull_mom-diff}, the effect of the interplay between instantaneous repulsion and retarded attraction is highly non-trivial and leads to new effects.
Since our calculation takes unbiasedly both ingredients into account and describes the experimental findings for the isotope effect in the pseudogap we conclude that such effects are not only relevant for cuprates but also for other strongly correlated systems in which the dominant mechanisms are of electronic origin but some selected phonon modes are still active.

\noindent
We wish to thank M.~Capone, S.~Ciuchi, E.~Gull, K.~Held, M.~Imada, A.~Millis and A.~Toschi for stimulating discussions. 
Calculations have been performed on the Vienna Scientific Cluster as well as on the machines of the Max-Planck Institute for Solid State Research, Stuttgart. 
G.S. acknowledges support from the FWF under ``Lise-Meitner'' Grant No. M1136.


\begin{thebibliography}{44}
\bibitem[{not()}]{note_el-dop}
\bibinfo{note}{Let us note that the {$e$}-doped side is further complicated by
  the presence of a coexistence region \cite{gull_mom-diff}}.
\end{thebibliography}

\end{document}